\documentclass[12pt,preprint]{aastex}
\usepackage{emulateapj5}

\def\fmag{\hbox{$.\!\!^m$}}
\newcommand{\mincir}{\raise
-2.truept\hbox{\rlap{\hbox{$\sim$}}\raise5.truept\hbox{$<$}\ }}
\newcommand{\magcir}{\raise
-2.truept\hbox{\rlap{\hbox{$\sim$}}\raise5.truept\hbox{$>$}\ }}

\newenvironment{inlinefigure}{%
\def\@captype{inlinefigure}%
\noindent\begin{minipage}{\linewidth}\begin{center}}
{\end{center}\end{minipage}\smallskip}
\makeatother

\begin{document}

\title{Morphological and Luminosity Content of Poor Galaxy Groups}

\author{Hrant M. Tovmassian\altaffilmark{1}, Manolis 
Plionis\altaffilmark{2,1} 
Heinz Andernach\altaffilmark{3}}

\altaffiltext{1}{Instituto Nacional de Astrof\'{\i}sica \'Optica y 
Electr\'onica, AP 51 y 216, 72000, Puebla, Pue, Mexico, email: 
hrant@inaoep.mx}

\altaffiltext{2}{National Observatory of Athens, 
Lofos Koufou, P.Penteli 152 36, Athens, Greece, email: 
mplionis@astro.noa.gr}

\altaffiltext{3}{Depto. de Astronom\'a, Univ.\ de Guanajuato, AP 144, 
Guanajuato, CP 36000, Mexico, email:heinz@astro.ugto.mx}

\begin{abstract}
We find that the fraction of early-type galaxies in poor 
groups (containing from 4 to 10 members) is a weakly increasing function 
of the number of the group members and is about two times higher 
than in a sample of isolated galaxies. We also find that the group
velocity dispersion increases weakly with the fraction of early-type 
galaxies. Early-type galaxies in poor groups are brighter in the 
near-infrared with respect
to isolated ones by $\Delta M_{K}\sim 
0\fmag75$, and to a lesser degree also in the blue ($\Delta M_{B} \sim 
0\fmag5$). We also find early-type galaxies in groups to be redder than those 
in the field. These findings suggest that the formation history for 
early-type galaxies in overdense regions is different from that of in 
underdense regions, and that their formation in groups is triggered by 
merging processes.
\end{abstract}

\keywords{galaxies: groups: general -- dynamics: galaxies --
morphology}

\section{Introduction}

It is known that most galaxies in the Universe occur in small groups 
(cf. Geller \& Huchra 1983; Tully 1987; Nolthenius \& White 1987; Fukugita et
al. 1998). 
In the dense and relatively low-velocity dispersion group environments one 
expects frequent galaxy interactions. Indeed, Tran et al. (2001) 
showed that a certain fraction of galaxies in evolved, X-ray luminous groups 
are significantly asymmetric which is evidence of galaxy interactions.
N-body simulations of isolated groups indicate that dynamical friction should 
also play an important role in the evolution of groups (Bode, Cohn, \& Lugger 
1993; Bode et al. 1994; Athanassoula, Makino, \& Bosma 1997). 

In such a scenario the groups are not in a dynamical equilibrium 
because of the high frequency of galaxy interactions. 
Interactions and merging of galaxies has been shown to play a major
role in the evolution of galaxy morphology (eg. Toomre \& Toomre 1972;
Schweizer \& Seitzer 1992).
The general belief is that early-type (E/S0) galaxies are formed by 
merging of spiral galaxies (Barnes \& Hernquist 1992; Mihos 1995). As a 
result of multiple merging, a massive, central galaxy can be formed. 
Obviously this phenomenon must be especially rapid in compact groups. 
For this reason a lot of efforts have been devoted to 
the study of dynamical evolution and morphological content of compact
groups (e.g. Hickson \& Mendes de Oliveira 1992, Mendes de Oliveira \& 
Hickson 1994; Tovmassian 2001, 2002; Kelm \& Focardi 2004; Lee et
al. 2004; Coziol, Brinks, \& Bravo-Alfaro 2004). 
Contrary to expectation, Shimada et al. (2000) found no statistical 
difference in the frequency of occurrence of emission-line galaxies between 
the Hickson compact groups (Hickson 1982; Hickson et al. 1992) and the field.
However, Kelm \& Focardi (2004) found that compact groups identified
in the Updated Zwicky catalogue of galaxies contain a higher fraction
of early-type galaxies with respect to the field. On the other hand,
Colbert, Mulchaey, \& Zabludoff (2001) have shown that early-type
galaxies in the field appear to have more shells and tidal features than 
those in groups, a fact that they attribute to merging events that occur also 
in the field, and whose signatures probably survive due to isolation of these 
galaxies.These results seem to be confirmed by Marcum, Aars, \& Fanelli 
(2004) although they have also identified a small fraction of isolated 
early-type galaxies that show no evidence of a merger history, and thus 
appear to be passively evolving primordial galaxies. 

Recently we have used the Ramella et al. (2002) 
UZC-SSRS2 group catalogue, based on the 
Updated Zwicky Catalogue (UZC; Falco et al. 1999) and the Southern 
Sky Redshift Survey (SSRS2; da Costa et al. 1998) 
to show that poor groups have a prolate spheroidal shape 
configuration with a mean intrinsic axial ratio ($\beta$) of $\langle 
\beta\rangle \approx0.3$ and standard deviation $\sigma_{\beta}\approx 0.15$ 
(Plionis, Basilakos, \& Tovmassian 2004, [hereafter PBT04]). 
As interaction are more likely in such elongated 
systems, poor groups are good laboratories to test galaxy formation theories 
and study their evolution.

In this paper we compare the morphological and luminosity content (in
the K and B bands) of the group galaxy members with those of isolated 
galaxies. 

\section{Observational Data}
We have used the poor groups of the UZC-SSRS2 catalogue (USGC, Ramella et 
al. 2002) as defined in PBT04, i.e., groups with galaxy membership $n_m$
(``richness`` in what follows) in the range $4\le n_m\le 10$ and 
radial velocity $cz \le 
5500$ km $s^{-1}$, since within this velocity limit the groups appear to have 
constant space density and therefore they are assumed to constitute a roughly 
volume-limited sample. As in PBT04, we divide groups in different
richness classes. All 
candidate fake groups (see PBT04) were excluded from our analysis. Furthermore 
we excluded those of our $n_m=4$ groups which Focardi \& Kelm (2002) 
identified as triplets (due to possible projection contamination), i.e., USGC 
U033, U039, U066, U070, U076, and  U127 (which correspond to FK 12, 14, 26, 
27, 29 and 35). As a starting point for our analysis we use
169 groups containing in total 932 galaxies, 
out of which 58, 59, 35 and 17 groups have a richness ($n_m$) of
4, 5-6, 7-8 and 9-10 galaxies, respectively.  

For comparison we used the catalogue of Isolated galaxies (IGs) compiled by 
Karachentseva, Lebedev \& Shcherbanovskij (1986). This catalogue is widely used 
as a source of isolated galaxies (e.g. Marcum et al. 2004; Varela et al. 
2004). It contains 1051 entries, among which there are 329 galaxies
with $cz \le 5500$ km $s^{-1}$. Note that this galaxy sample has the
same magnitude limit as our group galaxies ($m_{lim}\sim 15.5$). 
Two of the IGs, PGC 008220 and PGC 059971, were found in USGC groups 
and were deleted from the IG list.   

Galaxy morphological types for both poor group galaxies (GGs) and isolated
ones were taken from the NED (http://nedwww.ipac.caltech.edu). 
Since the galaxies we are dealing with are nearby and sufficiently
bright ($m_{\rm lim}\simeq 15.5$), we expect that the NED 
morphological classification into the two broad categories (early and 
late types) is very accurate.

We deduce absolute magnitudes in the $K$ band for two families of 
galaxies: (a) E/S0 galaxies and (b) spiral galaxies (Sa and later, including 
irregulars) for both the considered USGC group members and the
IGs. We used the 2{\sc MASS} 
$K_{\rm total}$ magnitudes (see Jarrett et al. 2000 and 
http://www.ipac.caltech.edu/2mass), corrected for the extinction in our 
Galaxy according to Schlegel, Finkbeiner, \& Davis (1998) as given in
NED. The quoted 2{\sc MASS} average magnitude 1$\sigma$ uncertainty is 
$\sim$0.04 for $m_k<11$ and $\sim$0.09 for $m_k\ge 11$.
The absolute $B$ magnitudes for all studied galaxies were extracted from LEDA
(Paturel et al. 1997, http://leda.univ-lyon1.fr) and their quoted 
individual 1$\sigma$ uncertainties are quite large, the average of
which is estimated to be $\sim 0.3$.

Note that in order to determine absolute magnitudes the radial velocities of
groups or individual galaxies were corrected for the peculiar velocity of the 
local group and a local velocity field that contains a Virgo-centric infall 
component and a bulk flow given by the expectations of linear theory (see 
Branchini, Plionis, \& Sciama 1996), assuming a Hubble constant of 
$H_0=70$ km s$^{-1}$ Mpc$^{-1}$. We excluded from our analysis groups
and isolated galaxies that have $cz \le 1000$ km s$^{-1}$, as 
their radial velocities may be significantly contaminated by peculiar 
velocities.

In order to investigate the mutual completeness of the GG and IG samples 
we employed the non-parametric Kolmogorov-Smirnov (KS) test 
and found that their redshift distributions may be considered 
(at a $\magcir$25\% level) as having  been drawn from the same 
parent population. This also implies that there is no systematic 
difference in their limiting magnitudes. 

\section{The fraction of early-type galaxies}
We determine the mean fraction of E/S0 galaxies, $f_{E/S0} (\equiv 
n(E/S0)/n_m$), over our group sample, as well as in the sample of
IGs. We use only those groups of which all members have known
morphological type, in total 102 groups spanning the whole original
sample richness range and which contain in total 583 galaxies 
(38, 27, 27 and 10 groups with $n_m=4$, 5-6, 7-8 and 9-10 members, 
respectively).

For the sample of 329 IGs we find $f_{E/S0}\sim 0.15$ while for the group 
galaxies we obtain a significantly higher mean fraction of 
$f_{E/S0}\sim 0.23$. This result is in accordance with
the ``morphology - density'' relation of Dressler (1981)
Kuntschner et al. (2002)  
also noticed that early-type galaxies are rare in low-density environments.
The higher relative number of E/S0 galaxies in groups favors the view
that they are formed by the interaction and merging of spiral galaxies.
The number of such galaxy encounters is expected to be even higher 
in high density environments, ie., in groups with large $n_m$.
To test this we estimate the mean fraction of early-type galaxies 
as a function of increasing group richness ($n_m$) and find: 
$f_{E/S0}\sim$ 0.21, 0.19, 0.29  and 0.26 for groups with $n_m=4$, 
5-6, 7-8, 9-10 members, respectively, suggesting a weak dependence of
$f_{E/S0}$ on the group richness.
However, it seems surprising that the relative number of E/S0's and possibly
their formation rate depends only weakly on the group richness and thus on 
the group mass. We investigate this issue further by correlating 
$f_{E/S0}$ with the velocity dispersion, $\sigma_v$, of the corresponding 
groups (Fig.~1), and we find that there is indeed a significant although weak 
correlation with the group velocity dispersion increasing with
$f_{E/S0}$ (in agreement with Zabludoff 2000). 
The Spearman rank correlation coefficient is 0.2 and 
the probability of random correlation is 0.045.
Meanwhile, such a 
correlation is not seen for groups of a given richness class, which confirms 
that the effect is due to an increase of the velocity dispersion with the 
group mass. The true underlying correlation should be even stronger than
the observed one since the dependence of the measured group velocity
dispersion on group orientation can weaken it. This effect
results from the fact that
elongated prolate-like groups orientated roughly along the line of sight 
will appear to have a higher velocity dispersion
while when seen orthogonal to the line of sight they
will appear to have smaller velocity dispersions.

\begin{inlinefigure}
\epsscale{1.04}	
\plotone{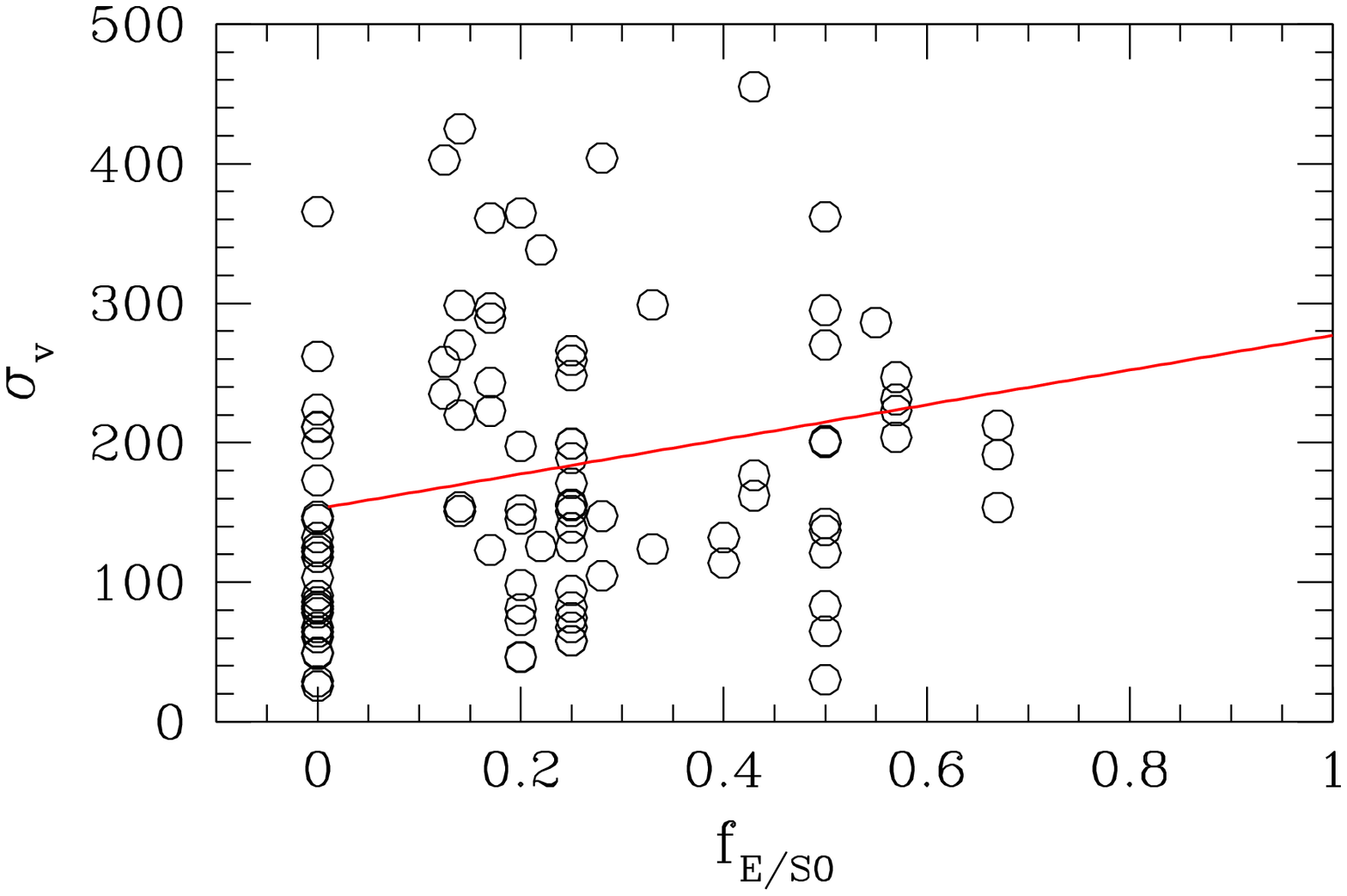}
\figcaption{The group velocity dispersion as a function of the
fraction of early-type galaxies for individual groups 
with $4\le n_m \le 10$. The continuous
line represents the best least-square fit, while the large star
symbols represent the average velocity dispersion in five $f_{E/SO}$ bins.}
\end{inlinefigure}

Another parameter that should play a significant role in the
quantification of the rate of galaxy encounters in groups is the group
crossing time, $\tau_c$, defined by:
$\tau_{c}=\overline{a}/\overline{\sigma}_v \;,$ 
with $\overline{a}$ the median true 
group major axis and $\overline{\sigma_v}$ the median true (deprojected) group 
velocity dispersion. 

The probability of an encounter in a group with a given number 
of members should be higher in groups with smaller $\tau_c$. However, 
effective interactions that could alter 
galaxy morphologies should happen in environments of relatively low velocity 
dispersion which implies a relatively large $\tau_c$ for a given group size. 
Note also that in high $\sigma_v$ systems that are also extremely elongated, 
effective galaxy interactions may happen at the apogees of the spatial 
configuration where the relative velocities of galaxies will be minimal and 
where the member galaxies will tend to accumulate. Clearly, the dynamics of 
galaxy mergers in non-spherical structures is a multi-parameter problem and 
not easily quantified.


We attempt to estimate the mean crossing time for groups of different
 richness ($n_m$) by taking into account the fact that they 
are prolate-like (Plionis et al. 2004) 
and that their member galaxies will probably have
mostly radial orbits (moving along the major axis of the prolate spheroid).
The {\em true} median $\sigma_v$  is estimated using the
quasi-spherical groups i.e. those with axial ratio close to unity, 
since due to orientation effects, 
the quasi-spherical groups should typically be
 those elongated groups seen end-on and therefore their velocity 
dispersion is closer to the true value. We find 
$\overline{\sigma}_v\simeq 170$ km/sec and 360 km/sec for the $n_m=4$ and
$n_m=9,10$ groups respectively.
In order to estimate the true size of groups we rely on a
Monte-Carlo simulation method, the details of which will be presented
elsewhere (Plionis \& Tovmassian 2004, in preparation). It is
based on searching for the intrinsic group major axis distribution 
that once folded through the intrinsic group axial ratio
distribution (determined in Plionis et al. 2004) and random
orientations with respect to the line of sight, will produce the projected
Monte-Carlo group shape parameters (axial ratio, minor and major axis) 
distributions which are in agreement with the corresponding observed
ones. 
We find $\overline{a}_{3D}\simeq 1.5\pm 0.3$ and $2\pm 0.5 \; h_{70}^{-1}$
Mpc, respectively for the $n_m=4$ and $n_m=9,10$ groups and therefore
the group crossing times are:
$\tau_{c}\sim 8.6$ and 5.8 Gyrs, respectively.



To test the effect of the different values of $\tau_c$ on the
galaxy encounter and merging rate, disentangling the possible effect 
of the different group orientations 
with respect to the line of sight, we determined the mean $f_{E/S0}$
in the most elongated chain-like groups (ie., those which are preferentially 
seen perpendicular to the line of sight) for two group categories: those with 
$\tau_c$ higher and smaller than the mean value of the corresponding group 
subsample. There are 36 groups with axial ratio $q<0.3$ 
(all of them with $n_m\le 6$) and with known morphology of all their members. 
We find 
$f_{E/S0} \simeq 0.25\pm0.055$ and $0.165\pm0.045$ for groups with $\tau_c$ 
smaller and larger than the corresponding group mean value, respectively.
Although the difference is not very significant it does imply, as expected,
that the crossing time is an important parameter 
in the quantification of the rate of galaxy interactions.

\section{Absolute magnitudes and colours of galaxies.} 
If E/S0 group galaxies are formed as the result of galaxy merging, one should 
expect them to be more luminous than isolated E/S0's. We estimated the mean 
$\langle M_{K}\rangle$ and $\langle M_{B}\rangle$ absolute magnitudes of E/S0 
and spiral galaxies (Sa and later) in our original group sample ($N=169$)
and isolated galaxies (see Table 1). The frequency distribution of the 
absolute magnitudes 
of the group and isolated early-type galaxies is shown in Fig. 2. We also 
derived the absolute magnitude difference ($\Delta M$) between GGs and IGs 
and found $\Delta M_K \approx0\fmag75$ and $\Delta M_B \approx0\fmag5$ for 
the E/S0's and $\Delta M_K \approx0\fmag05$ and $\Delta M_B 
\approx0\fmag02$ for the spirals (see Table 1). 

\begin{inlinefigure}
\epsscale{1.04}
\plotone{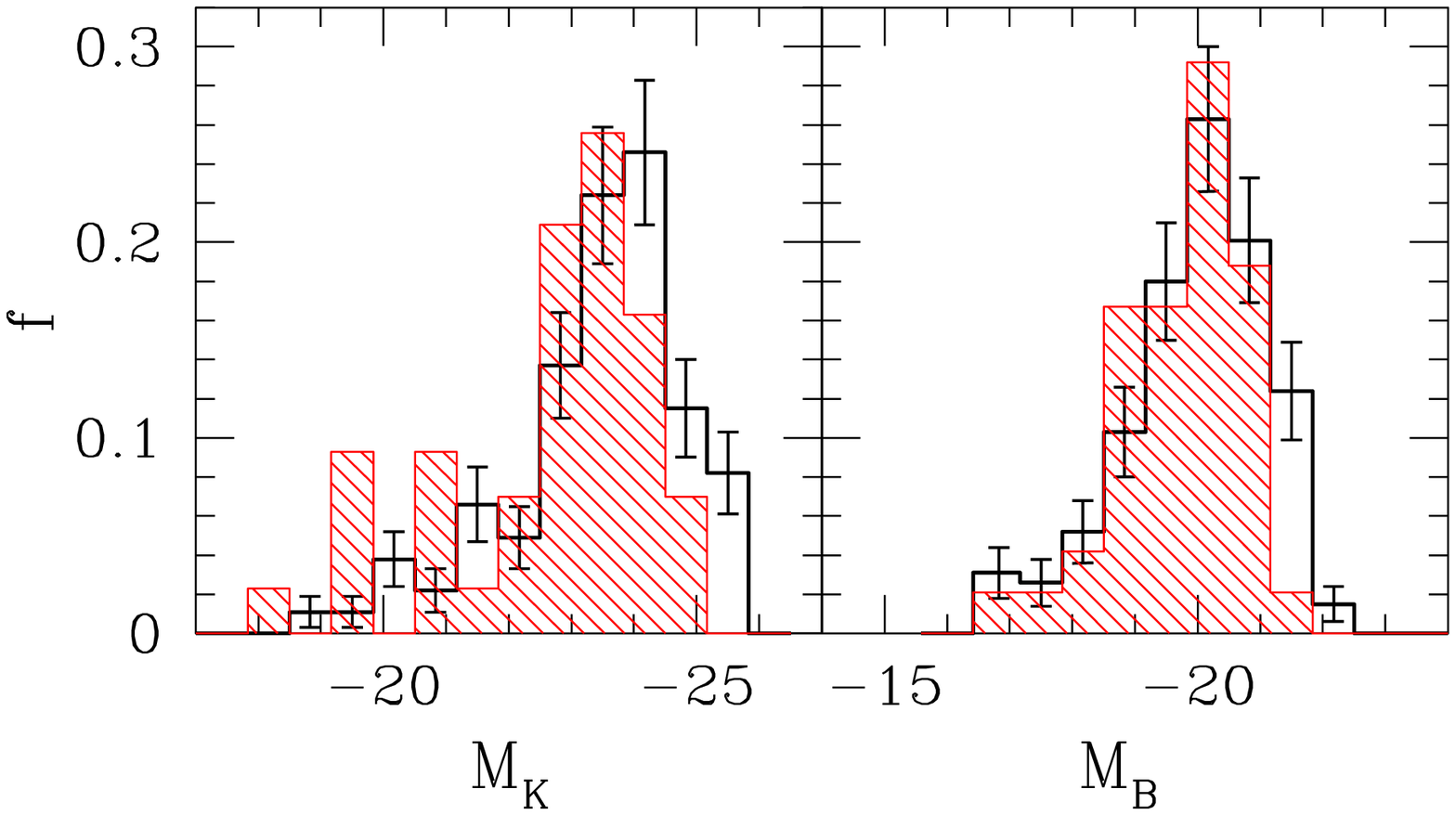}
\figcaption{The normalized frequency distribution of the $M_K$ (left) 
and $M_B$ (right) magnitudes for early-type galaxies in groups 
(line with error bars) and isolated (hatched histogram). 
The galaxy numbers involved are listed in Table 1.}
\end{inlinefigure}

The KS test shows that the GG and IG $K$-band
absolute magnitude distributions of 
early-type galaxies are significantly different, with the probability
${\cal P}\simeq 0.05$ of being drawn from the same population.
The corresponding $M_B$ distributions do not show a significant 
difference. Furthermore, no difference in absolute magnitude (neither for 
$M_K$ nor for $M_B$) is observed between group and isolated spirals. The KS 
test shows that their distributions have a probability of being drawn from 
the same population of ${\cal P}\simeq0.7$ and $0.95$ respectively for the 
$K$ and $B$ bands. We have also found that the values $\langle M_{B}\rangle$ 
and $\langle M_{K}\rangle$ of E/S0 galaxies do not depend on the group 
richness.
\begin{inlinefigure}
\epsscale{1.04}
\plotone{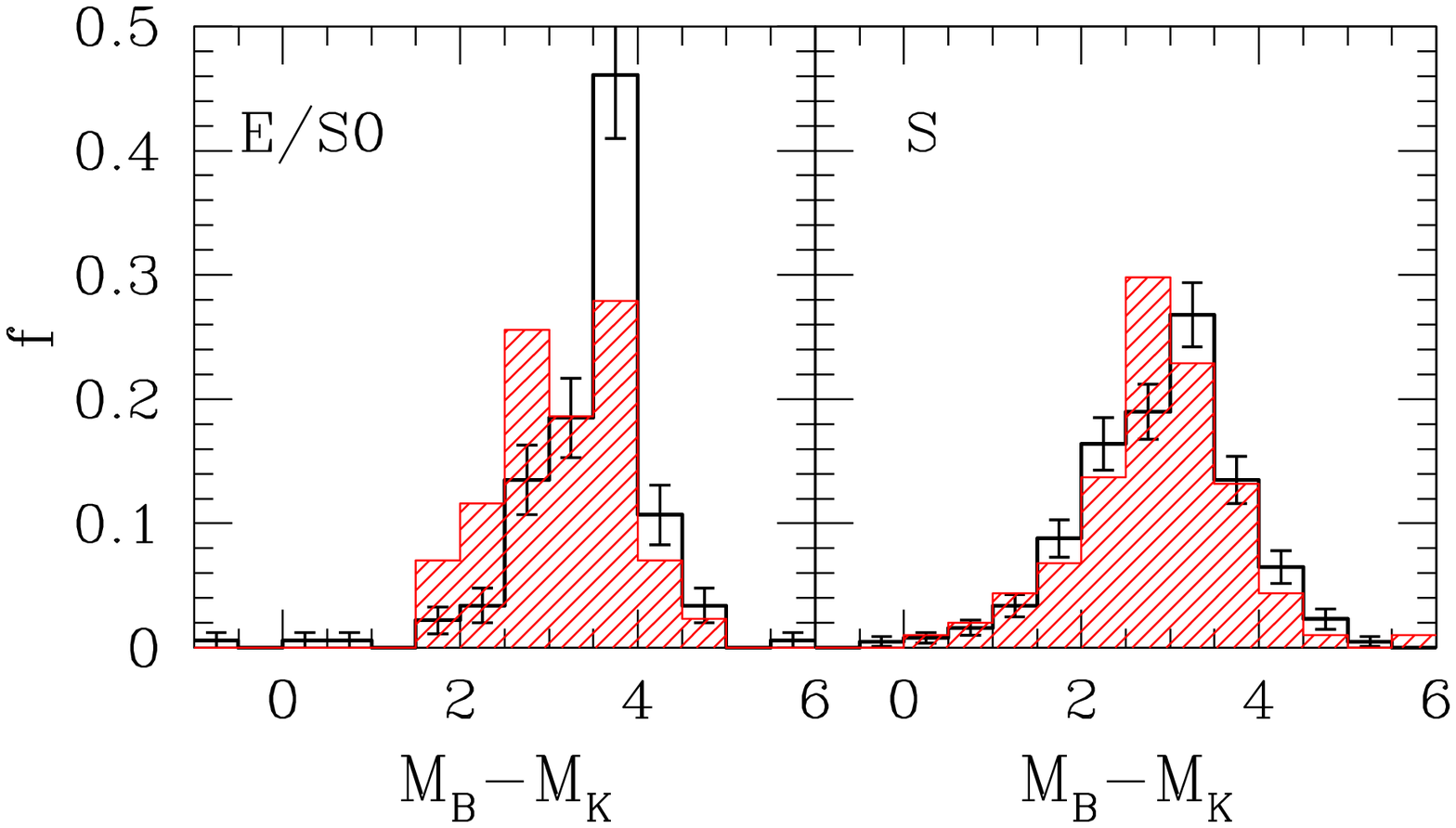}
\figcaption{Normalized frequency distribution of 
the $M_B-M_K$ colours of the group (line with error bars) and 
isolated (hatched) galaxies for early-types (left panel) and
spirals (right panel). The galaxy numbers involved are listed in Table 1.}
\end{inlinefigure}

We conclude that indeed there is strong evidence that the group E/S0 galaxies 
are brighter, especially in $K$ with respect to the IGs of either type, E/S0 
or spiral. The fact that absolute magnitudes of the group E/S0 galaxies are 
on average brighter than those of isolated galaxies by $\approx0\fmag75$ 
is consistent with the idea 
that E/S0 group galaxies are formed by merging of two 
spirals of about the same luminosity. The latter is in general agreement with 
the environmental dependence of the IR luminosity function found by Balogh et 
al. (2001).

Finally, we compare the $M_{B}-M_{K}$ colours of the considered subsamples. 
We find that the mean colour of the E/S0 galaxies in groups is redder by 
about $0\fmag30$ than that of the IGs, and the corresponding colour 
distributions are significantly different, having a KS probability of 
consistency of only ${\cal P}\simeq 0.006$. A redder colour of the group 
E/S0's may be due to shed of gas during merging. As expected, we find that 
both group and isolated spiral galaxies are bluer on average in comparison to 
the group and isolated E/S0 galaxies (see Fig.~3 and Table~1). Fig.~3 also 
shows the presence of a few relatively blue galaxies ($M_{B}-M_{K}\le 1$) 
among the group E/S0's for which weak starburst processes, induced by
interactions, could be responsible (Zepf, Whitmore, \& Levison 1991). 
 
\section {Conclusions}
We have found that the mean fraction of early-type galaxies, 
$f_{E/S0}$,
in poor USGC groups of galaxies (with $1000<cz<5500$ km s$^{-1}$ and
galaxy members $4\le n_m\le 10$) is $\sim 0.23 \pm 0.04$, which is
$\sim 65\%$ higher than that of the field galaxies. 
This fraction is a weakly increasing function of group richness and
thus of group mass. This trend is also confirmed from the
statistically significant, although weak, increase of 
$f_{E/S0}$ with group velocity dispersion.


We have also found that the mean near-infrared absolute magnitude of E/S0 
galaxies in our groups ($\langle M_K \rangle\approx-23\fmag4$) does not 
depend on the number of the group members and is $\sim 0\fmag75$ brighter 
than that of isolated E/S0 galaxies, while the mean blue absolute magnitude, 
$\langle M_B\rangle$, of E/S0 galaxies in these groups is brighter by 
$0\fmag5$ than the isolated ones. No such differences are found in the
corresponding spiral galaxy samples. We conclude that E/S0 galaxies in 
groups may be the result of merging of two galaxies of about the same 
luminosity. The $M_B-M_K$ colours of the group E/S0 galaxies are redder, by 
about $0\fmag30$ on average, than those of isolated E/S0 galaxies.

These results are in agreement with the paradigm in which the early-type 
galaxies, in relatively dense environments, are formed by merging. 
The redder $M_B-M_K$ colours of the group E/S0 galaxies in comparison to the 
isolated ones shows that the formation processes of the former somehow 
differs from that of isolated galaxies. An open question is still how the 
isolated E/S0 galaxies are formed. Various authors have found signs of 
disturbed morphologies in field E/S0's (eg. Colbert et al. 2001; Kuntschner 
et al. 2002; Marcum et al. 2004). Although the bright isolated E/S0 
galaxies may also be remnants of merged groups or pairs of galaxies, the 
isolated faint E/S0's probably are not (eg. Marcum et al. 2004); their
origin may be primordial and thus different from those in groups. 

\begin{acknowledgements}
This research has made use of the NASA/IPAC Extragalactic Database (NED) 
which is operated by the Jet Propulsion Laboratory, California Institute of 
Technology, under contract with the National Aeronautics and Space 
Administration and the Lyon-Meudon Extragalactik Database (LEDA, now 
Hyper-Leda) maintained at Observatoire de Lyon (see 
htpp://leda.univ-lyon.fr). MP acknowledges support by the Mexican 
CONACYT grant No 2002-C01-39679 and HA by CONACYT grant 40094-F. 
We thank the anonymous referee for comments that led to a clearer
presentation of our results.
\end{acknowledgements}

\newpage


\begin{table}[htb]
\caption[]{The mean absolute magnitudes $\langle M_K \rangle$ and $\langle
M_B \rangle$ of E/S0 and of spiral galaxies (Sa and later) in groups and in 
the field. In parenthesis the number of corresponding galaxies is given.}

\tabcolsep 3pt
\begin{tabular}{cccc|ccc} \\ \hline
  &   \multicolumn{3}{c}{E/S0} & \multicolumn{3}{c}{S} \\ 
      & $\langle M_K \rangle$  & $\langle M_B \rangle$ & $\Delta M$ & 
$\langle  
M_K\rangle$  & $\langle M_B \rangle$ & $\Delta M$ \\ 
\hline
 GG   & $-23.42\pm 1.39$ (183) & $-19.92\pm1.17$ (194) & -3.47$\pm 0.76$ & 
$-22.51\pm1.51$ (398) & $-19.47\pm1.36$ (500) & -2.89$\pm 0.89$ \\
 IG   & $-22.68\pm 1.64$ (43)  & $-19.44\pm1.32$ (48) & -3.16$\pm 0.72$ &  
$-22.54\pm 1.46$ (205) & $-19.44\pm1.40$ (269) & -2.82$\pm 0.86$\\  

\hline
\end{tabular}

\end{table}

\end{document}